\documentclass[12pt]{article}

\pdfoutput=1

\usepackage{graphicx}
\usepackage{dcolumn}
\usepackage{bm}
\usepackage{hyperref}
\usepackage{color}
\usepackage{amsmath}
\usepackage{amssymb}
\usepackage{amsfonts}
\usepackage{multirow}
\usepackage{cite}
\usepackage{slashed}

\usepackage[top=30truemm,bottom=30truemm,left=25truemm,right=25truemm]{geometry}

\begin{document}


\def\a{\alpha}
\def\b{\beta}
\def\c{\varepsilon}
\def\d{\delta}
\def\e{\epsilon}
\def\f{\phi}
\def\g{\gamma}
\def\h{\theta}
\def\k{\kappa}
\def\l{\lambda}
\def\m{\mu}
\def\n{\nu}
\def\p{\psi}
\def\q{\partial}
\def\r{\rho}
\def\s{\sigma}
\def\t{\tau}
\def\u{\upsilon}
\def\v{\varphi}
\def\w{\omega}
\def\x{\xi}
\def\y{\eta}
\def\z{\zeta}
\def\D{\Delta}
\def\G{\Gamma}
\def\H{\Theta}
\def\L{\Lambda}
\def\F{\Phi}
\def\P{\Psi}
\def\S{\Sigma}

\def\approxprop{%
  \def\p{%
    \setbox0=\vbox{\hbox{$\propto$}}%
    \ht0=0.6ex \box0 }%
  \def\s{%
    \vbox{\hbox{$\sim$}}%
  }%
  \mathrel{\raisebox{0.7ex}{%
      \mbox{$\underset{\s}{\p}$}%
    }}%
}

\def\o{\over}

\def\gmtm{(g-2)_\mu}

\def\gmte{(g-2)_e}

\newcommand{\gsim}{ \mathop{}_{\textstyle \sim}^{\textstyle >} }
\newcommand{\lsim}{ \mathop{}_{\textstyle \sim}^{\textstyle <} }
\newcommand{\vev}[1]{ \left\langle {#1} \right\rangle }
\newcommand{\bra}[1]{ \langle {#1} | }
\newcommand{\ket}[1]{ | {#1} \rangle }
\newcommand{\EV}{ {\rm eV} }
\newcommand{\KEV}{ {\rm keV} }
\newcommand{\MEV}{ {\rm MeV} }
\newcommand{\GEV}{ {\rm GeV} }
\newcommand{\TEV}{ {\rm TeV} }
\newcommand{\1}{\mbox{1}\hspace{-0.25em}\mbox{l}}
\newcommand{\headline}[1]{\noindent{\bf #1}}
\def\diag{\mathop{\rm diag}\nolimits}
\def\Spin{\mathop{\rm Spin}}
\def\SO{\mathop{\rm SO}}
\def\O{\mathop{\rm O}}
\def\SU{\mathop{\rm SU}}
\def\U{\mathop{\rm U}}
\def\Sp{\mathop{\rm Sp}}
\def\SL{\mathop{\rm SL}}
\def\tr{\mathop{\rm tr}}
\def\mpl{M_{\rm Pl}}

\def\etmiss{\slashed{E}_T}

\def\IJMP{Int.~J.~Mod.~Phys. }
\def\MPL{Mod.~Phys.~Lett. }
\def\NP{Nucl.~Phys. }
\def\PL{Phys.~Lett. }
\def\PR{Phys.~Rev. }
\def\PRL{Phys.~Rev.~Lett. }
\def\PTP{Prog.~Theor.~Phys. }
\def\ZP{Z.~Phys. }

\def\dd{\mathrm{d}}
\def\ff{\mathrm{f}}
\def\BH{{\rm BH}}
\def\inf{{\rm inf}}
\def\ev{{\rm evap}}
\def\eq{{\rm eq}}
\def\SM{{\rm sm}}
\def\Mpl{M_{\rm Pl}}
\def\GeV{{\rm GeV}}

\def\Msusy{m_{\rm stop}}
\newcommand{\Red}[1]{\textcolor{red}{#1}}
\newcommand{\TL}[1]{\textcolor{blue}{\bf TL: #1}}

\def\amuSUSY{a_\mu^{\rm SUSY}}
\def\aeSUSY{a_e^{\rm SUSY}}

\def\micromegas{{\tt MicrOmegas}}

\newcommand{\beq}{\begin{equation}}
\newcommand{\eeq}{\end{equation}}
\newcommand{\beqn}{\begin{eqnarray}}
\newcommand{\eeqn}{\end{eqnarray}}

 \def\be   {\begin{equation}}   \def\ee   {\end{equation}}
 \def\ba   {\begin{array}}      \def\ea   {\end{array}}
 \def\bea  {\begin{eqnarray}}   \def\eea  {\end{eqnarray}}
 \def\bean {\begin{eqnarray*}}  \def\eean {\end{eqnarray*}}
 \def\nn{\nonumber}
  \def\noin{\noindent}

\baselineskip 0.7cm
\begin{titlepage}

\vskip 2.35cm
\begin{center}
\vskip 2.35cm
~
\vskip 0.35cm

{\bf\Large Explanation of electron and muon $g-2$ anomalies 
\\ \vskip 1.5mm
in the MSSM}

\vskip 1.2cm
{\large Marcin Badziak and Kazuki Sakurai}
\vskip 0.6cm
{\it Institute of Theoretical Physics, Faculty of Physics, University of Warsaw, ul.~Pasteura 5, PL--02--093 Warsaw, Poland}\\
\vskip 2.2cm

\abstract{
The current experimental values of anomalous magnetic moments of muon and electron deviate from the Standard Model predictions by few standard deviations, which might be a hint of new physics. 
The sizes and signs of these deviations are different and opposite between the electron and muon,
which makes it difficult to explain both of these anomalies in a consistent model without introducing large flavour-violating effects.
It is shown that they can be simultaneously explained in the Minimal Supersymmetric Standard Model (MSSM)
by arranging the sizes of bino-slepton and chargino-sneutrino contributions differently between the electron and muon sectors.
The MSSM spectrum features very light selectrons and wino-like chargino, while they can evade LHC constraints due to degenerate spectra.   
}

\end{center}
\end{titlepage}

\setcounter{page}{2}

\section{Introduction}

The anomalous magnetic moments of leptons,  $g-2$, are some of the most precisely measured quantities that test the validity of the Standard Model (SM). Interestingly, current experimental values of $g-2$ for the electron and muon deviate from the state-of-the-art SM predictions which may be a hint of new physics.  

The biggest discrepancy between the SM prediction and the experimental is observed for $\gmtm$ which is at the level of $3.7\sigma$~\cite{Bennett:2006fi,Blum:2018mom}:
\begin{equation}
 \Delta a_\mu \equiv a_\mu^{\rm exp} - a_\mu^{\rm SM}= \left(2.74 \pm 0.73\right) \times 10^{-9} \,, 
\end{equation}
where $a_\mu \equiv ((g-2)_\mu)/2$. 

More recently, the updated value of the fine structure constant resulted in a new SM prediction for 
$\gmte$~\cite{Parker:2018vye}, which is $2.4\sigma$ below the value obtained by the measurement~\cite{Hanneke:2008tm}:
\begin{equation}
 \Delta a_e \equiv a_e^{\rm exp} - a_e^{\rm SM}= -\left(8.8 \pm 3.6\right) \times 10^{-13} \,.
\end{equation}

The long-lasting $\gmtm$ anomaly has been exhaustively studied over the years and the most interesting new physics scenarios explaining it include low-energy supersymmetry (SUSY)~\cite{Moroi:1995yh,Martin:2001st}, light pseudoscalar~\cite{Jackiw:1972jz,Leveille:1977rc} and dark photon~\cite{Pospelov:2008zw}. 
Providing a common explanation for both $\gmte$ and $\gmtm$ anomalies is more challenging. This is because in generic models without flavour violation new contribution to $g-2$ of a given lepton is proportional to the mass squared of that lepton. 
Namely, if new physics is flavour blind, its contributions to the anomalous magnetic moments, 
$a_e^{\rm NP}$ and $a_\mu^{\rm NP}$, are related in general as
\beq
\frac{m_\mu^2}{m_e^2} \frac{a^{\rm NP}_e}{a^{\rm NP}_\mu}
\,\sim\, 1 \,.
\label{eq:ratio1}
\eeq
This is in a sharp contrast with the above experimental observation, which indicates 
a different relation 
\beq
\frac{m_\mu^2}{m_e^2} \frac{\Delta a_e}{\Delta a_\mu}
\,\sim\, -14 \,.
\label{eq:ratio2}
\eeq
The program of explaining both $\gmte$ and $\gmtm$ anomalies by the same new physics faces two problems.
Firstly, such a new physics model has to give an order of magnitude larger contribution to the $\gmte$ 
than naively expected from the contribution to the $\gmtm$.
Second, the model must give the contributions to $\gmte$ and $\gmtm$ with the opposite signs.

There has already been several attempts to explain the experimental results for muon and electron $g-2$. It was argued in ref.~\cite{Crivellin:2018qmi} that this requires new sources of flavour violation, see also ref.~\cite{Giudice:2012ms, Fayet:2007ua}. 
One solution is to introduce a single CP-even scalar with sub-GeV mass that couples differently to muons and electrons~\cite{Davoudiasl:2018fbb}. In ref.~\cite{Liu:2018xkx} both anomalies were explained by introducing a light complex scalar that is charged under a global U(1) under which the electron is  also charged but muon not. 
Axion-like particles with lepton-flavour violating couplings explaining measured values of $\gmte$ and $\gmtm$ were investigated in ref.~\cite{Bauer:2019gfk}. 
Lepton-flavour violating couplings were considered also in a model with additional Higgs doublet \cite{Han:2018znu}.
It was also shown that appropriately large contribution to the $\gmte$ can be obtained 
for light sleptons in Minimal Supersymmetric Standard Model (MSSM) 
if flavour violating off-diagonal elements are introduced in the slepton mass matrices~\cite{Dutta:2018fge}.

In the present work, we will show that both anomalies can be simultaneously explained in the MSSM without introducing explicit flavour mixing. 
We make use of the fact that there are two one-loop diagrams that may give significant contributions to $g-2$ in the MSSM. 
These contributions may have a different sign and they have different scaling behaviour for increasing slepton masses. These features allow the total SUSY contributions to $\gmte$ and $\gmtm$ to have a different sign and correct magnitude to explain the experimental data if smuons (or at least the right-handed one) are heavier than selectrons.
Fitting simultaneously $\gmte$ and $\gmtm$ leads to quite sharp prediction for the electroweak part of the MSSM spectrum with some sleptons and wino-like chargino with masses not far above the LEP bound of about 100~GeV~\cite{LEP} but consistent with the LHC constraints.

The rest of the paper is organized as follows. In Section~\ref{sec:g2el} we briefly review dominant contributions to the lepton $g-2$ in the MSSM and qualitatively discuss properties of the MSSM spectrum leading to a good fit to the measurements of electron and muon $g-2$. In the following sections we discuss two scenarios in more detail paying a particular attention to the LHC constraints. In Section~\ref{sec:g2elmu} we discuss a case with the right-handed smuon much heavier than the rest of the sleptons of the first two generations, while in Section~\ref{sec:Higgsmed} we discuss a case motivated by the Higgs mediated SUSY breaking scenario with both left- and right-handed smuons heavier than selectrons.
We reserve Section~\ref{sec:concl} for conclusions and final remarks.

\section{Supersymmetric contribution to lepton magnetic moment}
\label{sec:g2el}

In the MSSM there are two types of the leading one-loop contributions to the anomalous magnetic moment of a lepton $l$;
the contribution coming from a chargino-sneutrino loop, $a^{\chi^{\pm}}_{l}$,
and the one coming from a bino-slepton loop, $a^{\chi^0}_l$.
They are approximately given by \cite{Moroi:1995yh,Martin:2001st}
\begin{eqnarray}
\label{amuC}
a^{\chi^{\pm}}_{l} &\approx&
\frac{\alpha \, m^2_l \, \mu\,M_{2} \tan\beta}
{4\pi \sin^2\theta_W \, m_{\tilde{\nu}_{l}}^{2}}
\left( \frac{f_{\chi^{\pm}}(M_{2}^2/m_{\tilde{\nu}_{l}}^2)
-f_{\chi^{\pm}}(\mu^2/m_{\tilde{\nu}_{l}}^2)}{M_2^2-\mu^2} \right) \, ,
\\
\label{amuN}
a^{\chi^0}_l &\approx&
\frac{\alpha \, m^2_l \, \,M_{1}(\mu \tan\beta-A_l)}
{4\pi \cos^2\theta_W \, (m_{\tilde{l}_R}^2 - m_{\tilde{l}_L}^2)}
\left(\frac{f_{\chi^0}(M^2_1/m_{\tilde{l}_R}^2)}{m_{\tilde{l}_R}^2} 
- \frac{f_{\chi^0}(M^2_1/m_{\tilde{l}_L}^2)}{m_{\tilde{l}_L}^2}\right)\,,
\end{eqnarray}
where $m_{\tilde{l}_L}$ and $m_{\tilde{l}_R}$ are slepton 
masses, and the loop functions are given by
\begin{eqnarray}\label{loopf1}
f_{\chi^{\pm}}(x) &=& \frac{x^2 - 4x + 3 + 2\ln x}{(1-x)^3}~,
\qquad ~f_{\chi^{\pm}}(1)=-2/3\, ,  \\
\label{loopf2}
f_{\chi^0}(x) &=& \frac{ x^2 -1- 2x\ln x}{(1-x)^3}\,,
\qquad\qquad f_{\chi^0}(1) = -1/3\, . 
\end{eqnarray}
In our numerical analysis, we compute the total SUSY contribution, $a_l^{\rm SUSY}$, at one-loop level following the calculation of ref.~\cite{Martin:2001st}.\footnote{
We also cross-checked our results with a routine for calculation of $\amuSUSY$ in \micromegas~code~\cite{Belanger:2006is} and found a good agreement.} 

The muon $g-2$ anomaly has been exhaustively studied in the MSSM for many years, see e.g.~\cite{Martin:2001st,Baer:2001kn,Stockinger:2006zn,Endo:2013lva,Badziak:2014kea,Kowalska:2015zja,Lindner:2016bgg}.  It is well known that the current central value of muon $g-2$ can be explained with either chargino-sneutrino~\cite{Martin:2001st} or bino-smuon~\cite{Endo:2013lva} contribution even if one of these contributions is negligible. However, obtaining the observed deviation in the electron $g-2$ is more challenging because it requires an order of magnitude larger SUSY contribution after taking into account the universality scaling factor $m_e^2/m_\mu^2$. 
We found that it is difficult to explain the current central value of electron $g-2$ with chargino-sneutrino contribution alone because it requires the left-handed selectron and both the wino and higgsinos to be very light, which are subject to tight collider constrains.
It is much easier to explain the current central value of electron if bino contribution is non-negligible.

Since the measurement of muon $g-2$ shows even larger deviation from the SM prediction an interesting question appears whether both $\gmte$ and $\gmtm$ can be explained simultaneously in the MSSM. This is not possible in a universal case of the same soft masses for smuons and selectrons because this predicts the ratio of SUSY contributions to $\gmte$ and $\gmtm$ to be $m_e^2/m_\mu^2$. 
This means not only the magnitude but also signs are wrong.
Therefore, in order to explain the current central values 
there must be splitting between smuon and selectron masses. 
Moreover, in order to obtain the opposite signs for $a_e^{\rm SUSY}$ and $a_\mu^{\rm SUSY}$ 
the SUSY contributions to $\gmte$ and $\gmtm$ 
 must be dominated by different diagrams with different signs.  
Interestingly, this is possible in the MSSM because the sign of the chargino-sneutrino is given by sign$(\mu M_2)$, while that of the bino-slepton contribution is sign$(\mu M_1)$.  
Therefore if $M_1 M_2<0$
and the SUSY contributions are dominated by different diagrams, 
$\aeSUSY$ and $\amuSUSY$ can have different signs.

We found that explaining electron $g-2$ with a large negative chargino-sneutrino requires left-handed selectron and both charginos to be close to the LEP bound ($m \gtrsim 100$~GeV). 
In this part of the parameter space, the bino-smuon contribution is too small to explain the muon $g-2$ 
even for light smuons just above the LEP bounds.
Thus, in what follows, we assume $\aeSUSY$ is dominated by the bino-selectron contribution and take $\mu M_1<0$ to make this contribution negative. 
We also choose $\mu M_2>0$ to make the chargino-sneutrino contribution positive 
which is supposed to dominate $\amuSUSY$. 
We identified two patterns of non-universalites in the MSSM spectrum that allow for negative contribution to $\gmte$ and positive contribution to $\gmtm$ with the correct magnitude to explain the experimental values. 
In each case selectrons and wino-like chargino and neutralino must be very light with masses not far above the LEP bound but fulfilling LHC constraints. 
In order to explain the current experimental value of $\gmte$ one also needs a large left-right mixing term,
$m_e (\mu \tan\beta-A_e)$, in the selectron mass matrix to enhance the bino-selectron contribution. 
This prefers relatively large $\mu$ unless $A_e$ is very large. In the following analysis we assume for simplicity vanishing $A$-terms. 
We discuss other features of the MSSM spectrum that lead to simultaneous explanation of both anomalies in the following sections.

\section{Heavy right-handed smuon}
\label{sec:g2elmu}

In this section, we pursuit an idea that
$\gmtm$ is explained by the chargino-sneutrino contribution by suppressing (opposite-sign) bino-smuon contribution by making the right-handed smuon heavy, while $\gmte$ is explained by the large negative bino-selectron contribution
assisted by relatively large left-right mixing term, $m_e (\mu \tan\beta-A_e)$.
In order to conveniently show the effect of the bino-slepton and chargino-sneutrino contributions,  
we introduce
\beqn
R_l^{\chi^\pm/\chi^0} = \frac{2 a_l^{\chi^\pm/\chi^0} -  \Delta a_l}{2 \sigma_l} \,,
\eeqn
such that
\beqn
R_l^{\rm SUSY} = \frac{a_l^{\rm SUSY} - \Delta a_l}{\sigma_l} = R_l^{\chi^\pm} + R_l^{\chi^0}  \,.
\eeqn
Namely, $R^{\rm SUSY}_l$ represents the standard deviations between the model and experiment,
and $R_l^{\chi^0}$ and $R_l^{\chi^\pm}$ are the bino-slepton and chargino-sneutrino contributions to $R^{\rm SUSY}_l$, respectively.
Note that in the limit where the SUSY particles are heavy and decoupled, 
$R^{\rm SUSY}_l$ approaches to the currently observed deviation, i.e.~$R^{\rm SUSY}_\mu \sim 3.8$ and $R^{\rm SUSY}_e \sim -2.4$, while $R^{\chi^0/\chi^\pm}_l$ approaches a half of that value.

\begin{figure}[t!]
\centering
\includegraphics[scale=0.67]{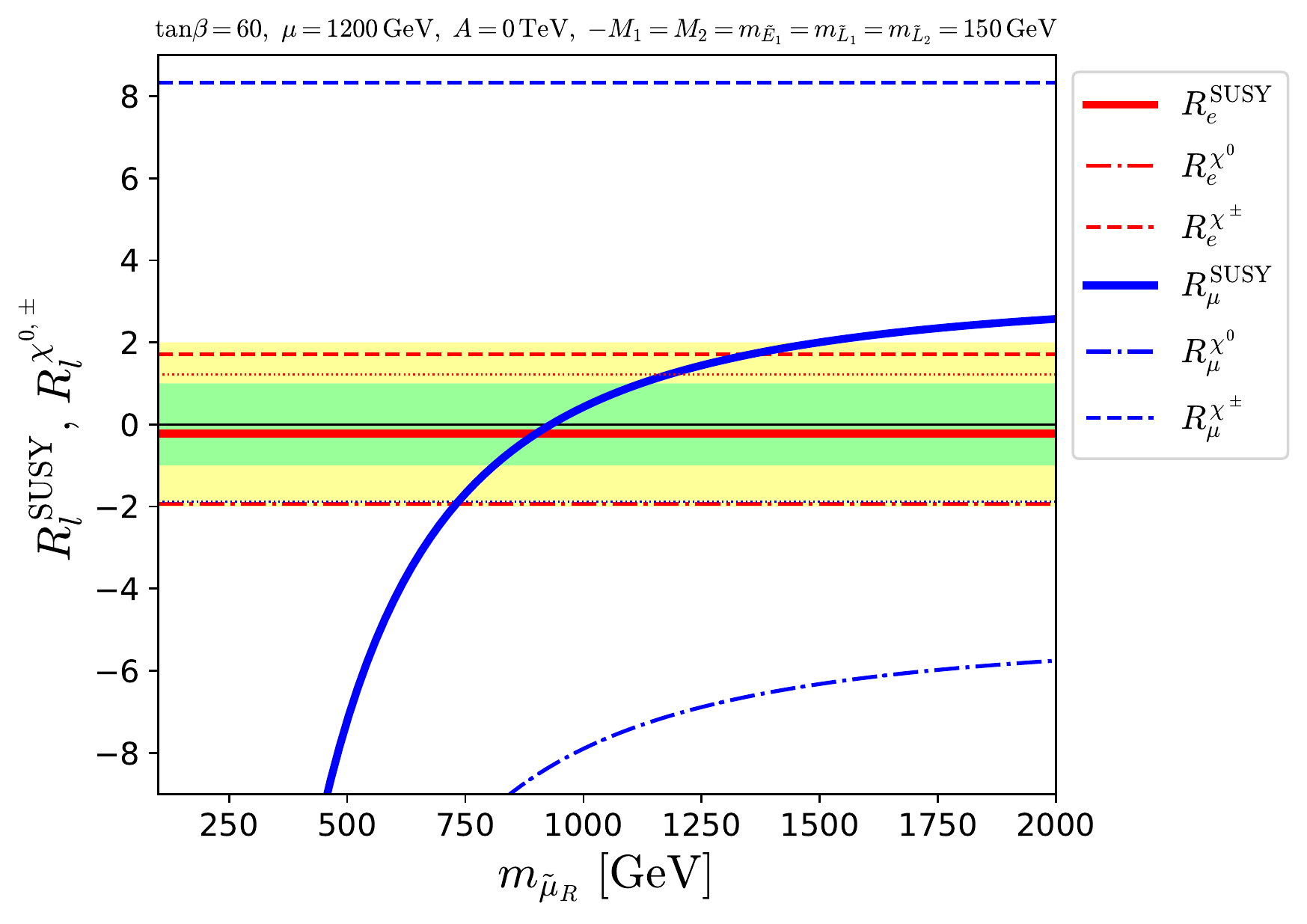} 
\caption{\small $R^{\rm SUSY}_l$ (solid), $R^{\chi^\pm}_l$ (dashed) and $R^{\chi^0}_l$ (dashed-dotted) for electron (red) and muon (blue) as a function of $m_{\tilde{\mu}_R}$. Very thin dotted lines around 1 (red for electron) and -2 (blue for muon) correspond to $- \Delta a_l/(2 \sigma_l)$, which $R^{\chi^\pm}_l$ and $R^{\chi^0}_l$ approach in the decoupling limit.  }
\label{fig:1D_muR}
\end{figure}

\begin{figure}[h!]
\includegraphics[scale=0.65]{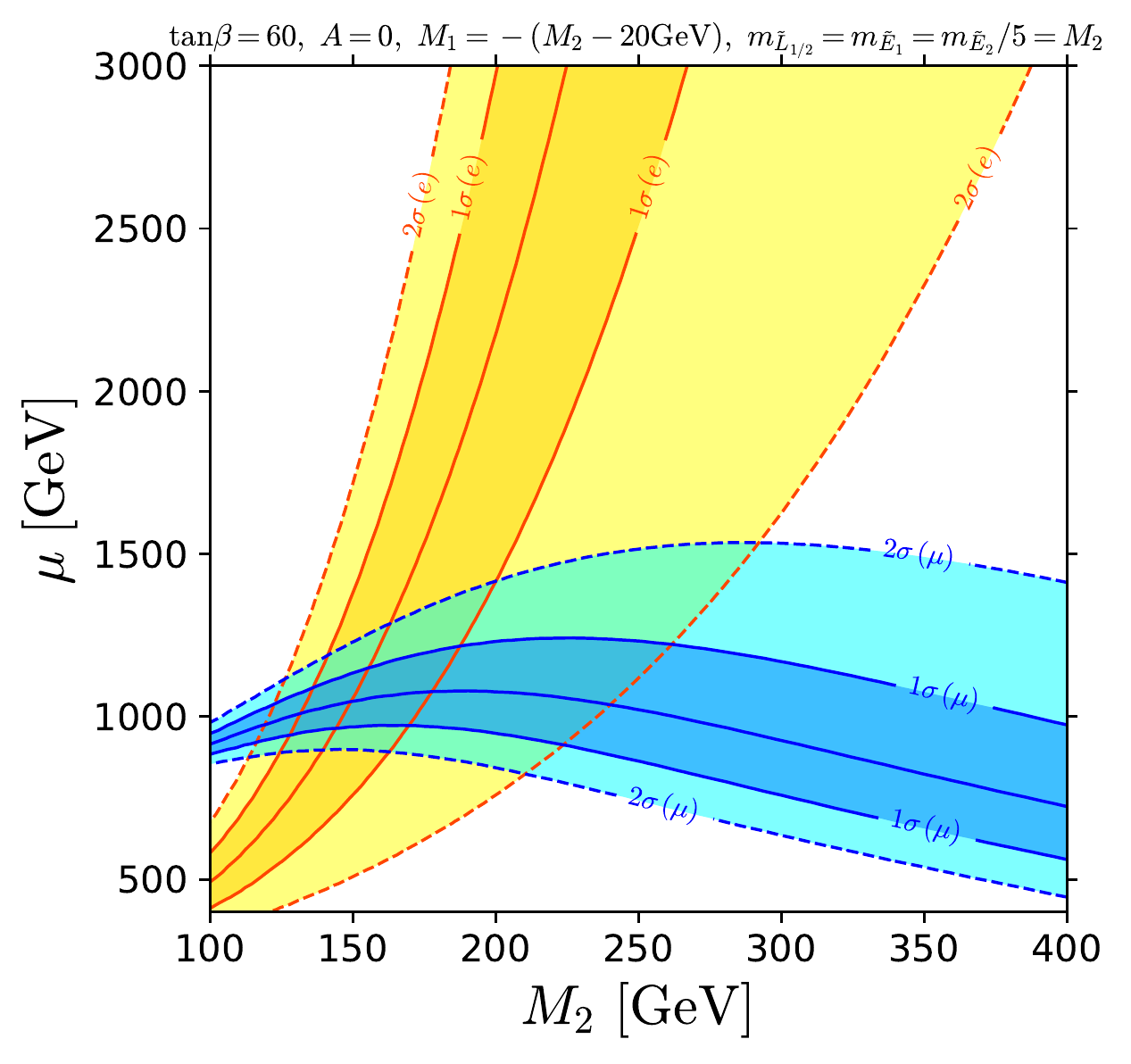}
\includegraphics[scale=0.65]{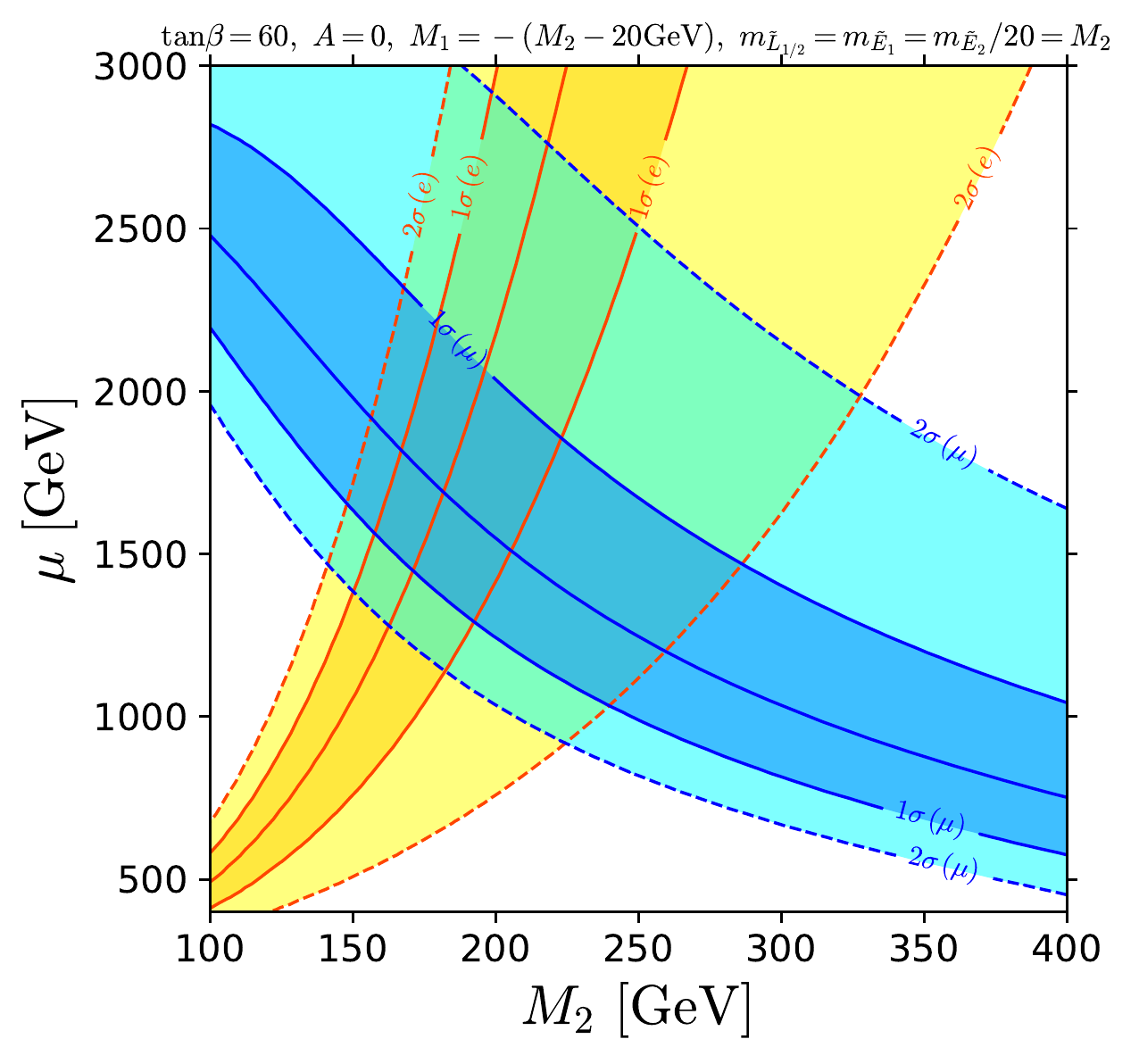}
\caption{\small
Contours of SUSY contribution to electron (yellow) and muon (blue) $g-2$ in the plane of $\mu$ and $M_2=m_{\tilde E_1} = m_{\tilde L_1} = m_{\tilde L_2}$ for $\tan\beta=60$ and $M_1=-(M_2-20\,\GeV)$. 
In the left (right) plot $m_{\tilde{\mu}_R}$ is 5 (20) times larger than $M_2$ and the other sleptons.}
\label{fig:g2_mu_mL}
\end{figure}

In fig.~\ref{fig:1D_muR} we present $R_l$'s as a function of the right-handed smuon mass, $m_{\widetilde \mu_R}$. The other parameters are fixed in such a way that the SUSY prediction of $\gmte$ is in a good agreement with the experimental value. 
We see that for very low $m_{\widetilde \mu_R}$, the deviation of $\gmtm$ from the experimental value
is large and negative, where the SUSY contribution to $\gmtm$ has a wrong sign.
This is because in this region the bino contribution dominates $\amuSUSY$,
which is proportional to $M_1 \mu < 0$. 
When increasing the $m_{\widetilde \mu_R}$ the absolute value of the bino contribution to $\gmtm$ gets smaller and at some point becomes subdominant with respect to the chargino contribution, driving $\amuSUSY$ positive. For the range of $800 \gtrsim m_{\widetilde \mu_R}/{\rm GeV} \gtrsim 1100$, $\gmtm$ is also within $1\sigma$ from the central values and there is even a point in the parameter space for which $\gmte$ and $\gmtm$ sit simultaneously at experimental central values.

\begin{figure}[t!]
\includegraphics[scale=0.67]{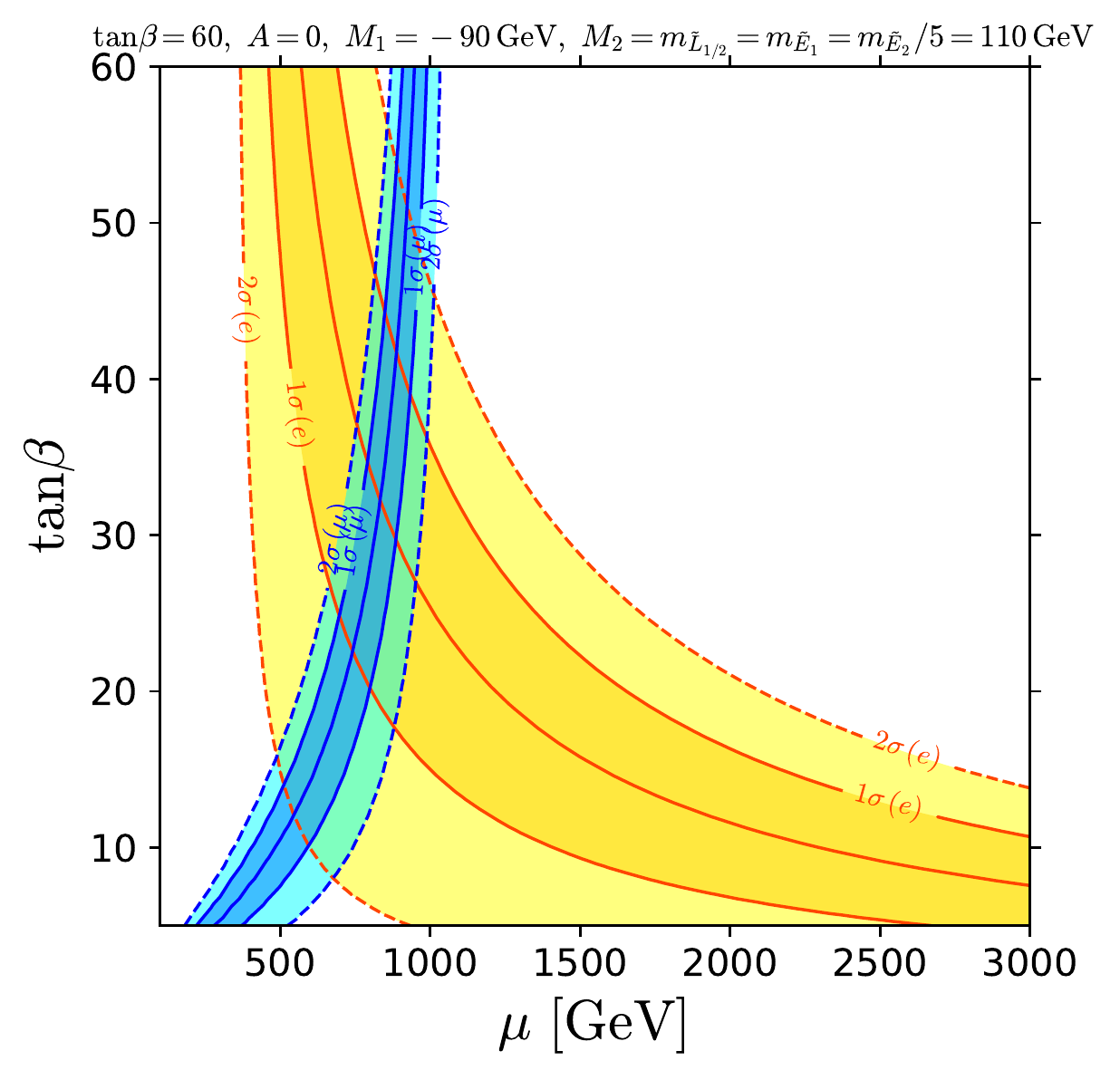} 
\includegraphics[scale=0.67]{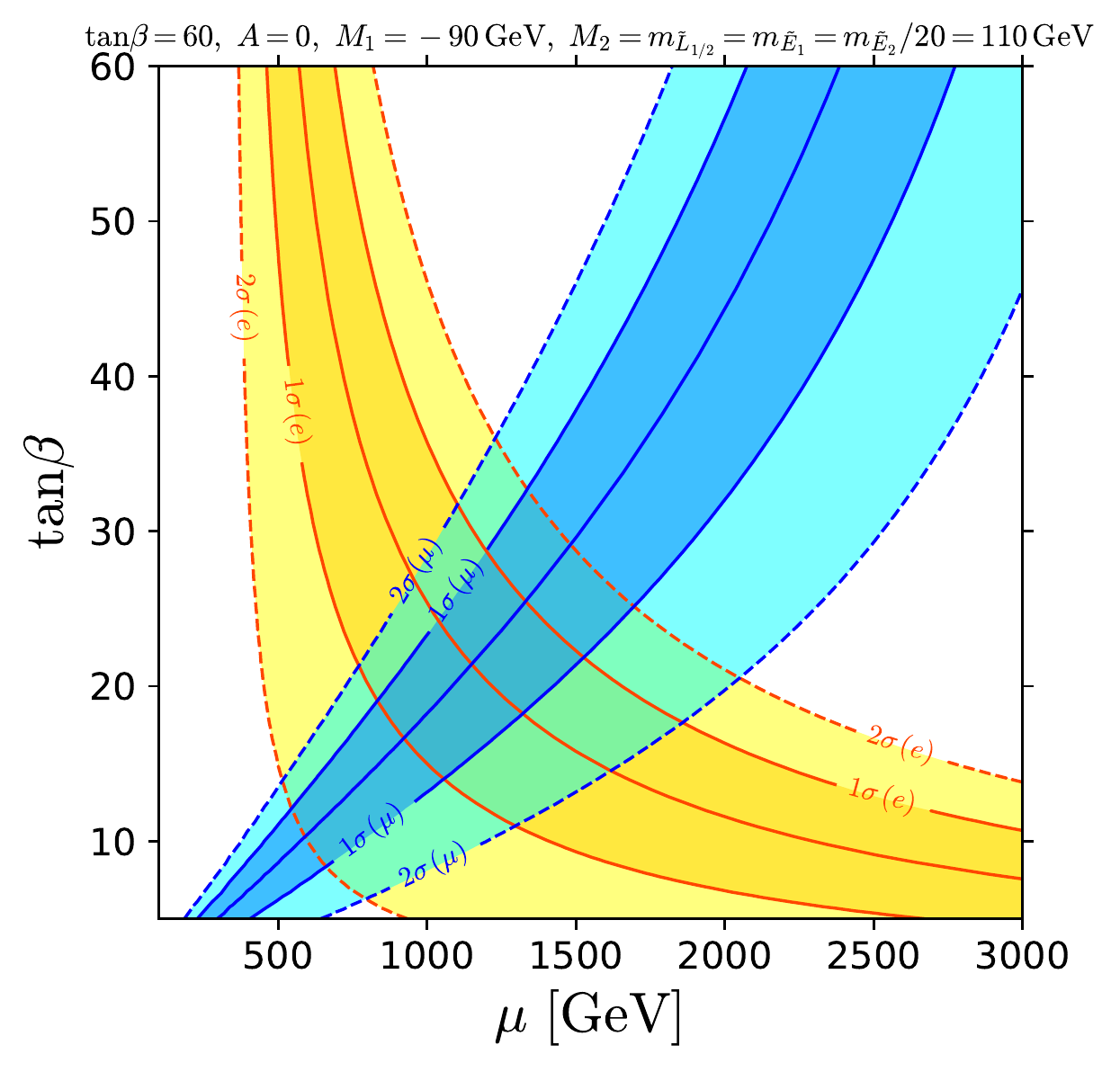}
\caption{\small The same as in  Fig.~\ref{fig:g2_mu_mL} but in the plane $\tan\beta$ vs. $\mu$ for $m_{\tilde E_1} = m_{\tilde L_1} = m_{\tilde L_2}=M_2=110$~GeV and $M_1=-90$~GeV. }
\label{fig:g2_mu_tan}
\end{figure}

In order to simultaneously explain $\gmte$ and $\gmtm$ selectrons must be rather light and
a preferred range of parameters strongly depends on $m_{\widetilde \mu_R}$, the chargino masses and the rest of the slepton masses. 
For smaller $m_{\widetilde \mu_R}$, the wrong-sign bino-smuon contribution increases which must be compensated by increasing the chargino-sneutrino contribution that has the correct sign. 
Since $a^{\chi^0}_l \approxprop \mu$ while $a^{\chi^{\pm}}_{l} \approxprop 1/\mu$ in the limit of large $\mu$, the chargino-sneutrino contribution becomes relatively more important as $\mu$ decreases. 
However, for smaller $\mu$ selectrons must be very light to account for the observed value of $\gmte$. 
It can be seen in the left panel of Fig.~\ref{fig:g2_mu_mL} that for the right-handed smuon mass five times larger than the mass of the other sleptons, the latter must be below 180 GeV in order to explain both $\gmte$ and $\gmtm$ at $1\sigma$ level, while the preferred higgsino mass is around 1~TeV.
On the other hand, in the limit of a very heavy right-handed smuon, cf.~the right panel of Fig.~\ref{fig:g2_mu_mL}, the other slepton masses can be as large as 220 (330) GeV to fit both $\gmte$ and $\gmtm$ at the $1\sigma$ ($2\sigma$) level. 
The preferred higgsino mass in this case is around $1.5 - 2$ TeV.

In the analysis so far we fixed $\tan\beta=60$. The $\gmte$ can be explained also for smaller values of $\tan\beta$ but at the cost of larger $\mu$. Since smaller $\tan\beta$ and larger $\mu$ makes $\amuSUSY$ smaller, fitting $\gmte$ and $\gmtm$ simultaneously implies a lower bound on $\tan\beta$. 
In Fig.~\ref{fig:g2_mu_tan} we show  contours of $\aeSUSY$ and $\amuSUSY$ in the plane of $\mu$ and $\tan\beta$ fixing the wino, selectron and left-handed smuon masses at 110 GeV, i.e.~just above the LEP bound. We see that $\tan\beta$ as small as 15 is sufficient to explain $\gmte$ and $\gmtm$ at $1\sigma$ level in the limit of decoupled right-handed smuon. 
For lighter right-handed smuon the lower bound on $\tan\beta$ is more stringent because the bino-smuon contribution partially cancel the chargino-sneutrino one. 
For the right-handed smuon five times heavier the right-handed selectron $\tan\beta$ must be at least 20 
to fit both $\gmte$ and $\gmtm$ within the $1\sigma$ level. 
Allowing $2\sigma$ deviations $\tan\beta$ of about 10 is also possible.

\subsection{LHC constraints}

We have just seen that in order to simultaneously fit $\gmte$ and $\gmtm$ very light sleptons and chargino  are necessary. Such light sparticles are generically strongly constrained by the LHC data. 
However, unlike coloured particles, 
the limits on electroweakly interacting particles are not very strong,
especially when the spectrum is compressed.
In this subsection, we provide an example spectrum that can fit both $g-2$ anomalies 
and is allowed by the current LHC data.
Our benchmark point, BP-1, is defined by the following parameters:
\beqn
M_1 = -180\,{\rm GeV},~~~M_2 = 170\,{\rm GeV},~~~~~~~~~~~~~~~
\nonumber \\
~~~~~~~~~~~~m_{\tilde E_1} = m_{\tilde L_1} = m_{\tilde L_2} = 200\,{\rm GeV},
~~~ 
m_{\tilde E_2} = 2000\,{\rm GeV},~~~ 
\nonumber \\
\mu = 1700\,{\rm GeV},~~~
A = 0,~~~
\tan \beta = 60.~~~~~~~~~~~~~~
\eeqn
With this set of parameters, the SUSY contributions to $(g-2)_e$ and $(g-2)_\mu$ are:
\begin{align}
\aeSUSY &= -\,6.50 \cdot 10^{-13},  &  R^{\rm SUSY}_e &=0.64,  &    (a_e^{\tilde \chi_1^0}, a_e^{\tilde \chi_1^\pm}) &= (-7.478, ~0.981) \cdot 10^{-13},             \\
\amuSUSY &= ~~\,2.36 \cdot 10^{-9},  &  R^{\rm SUSY}_\mu &= -0.52, &
(a_\mu^{\tilde \chi_1^0}, a_\mu^{\tilde \chi_1^\pm}) &= (-1.829,~ 4.193) \cdot 10^{-9},
\end{align}
which are within the 1-$\sigma$ bands of the measurements. 
The masses of the physical states are calculated using {\tt Suspect} \cite{Djouadi:2002ze}
and shown in Table \ref{tb:mass}. 
We also show the relevant branching ratios, calculated using {\tt SDecay} \cite{Muhlleitner:2003vg}, and cross-sections, calculated using {\tt Prospino 2} \cite{Beenakker:1999xh} at the NLO accuracy,
in Table~\ref{tb:xs_br}.

\begin{table}[t!]
\begin{center}
\begin{tabular}{cc}
$\tilde \chi_1^0$ & 175.7 \\
$\tilde \chi_2^0$ & 179.7 \\
$\tilde \chi_1^\pm$ & 179.7 \\
\end{tabular}
~~
\begin{tabular}{cc}
$\tilde \nu_e$ & 189.6 \\
$\tilde e_L$ & 205.1 \\
$\tilde e_R$ & 204.9 \\
\end{tabular}
~~
\begin{tabular}{cc}
$\tilde \nu_\mu$ & 189.6 \\
$\tilde \mu_L$ & 205.1 \\
$\tilde \mu_R$ & 2000 \\
\end{tabular}
~~
\begin{tabular}{cc}
$\tilde \chi_3^0$ & 1702 \\
$\tilde \chi_4^0$ & 1702 \\
$\tilde \chi_2^\pm$ & 1703 \\
\end{tabular}
\caption{\small \label{tb:mass} \small
Physical masses in GeV at the benchmark model point BP-1.}
\vspace{1.cm}
\begin{tabular}{|r|c|}
\hline
mode & $\sigma$\,[fb] \\ 
\hline
$\tilde \chi_1^\pm \tilde \chi_2^0$ & $2.44 \cdot 10^3$ \\
$\chi_1^+ \tilde \chi_1^-$ & $ 1.21 \cdot 10^3$ \\
\hline
$\tilde e^+_R \tilde e^-_R$ &     7.41     \\
$\tilde e^+_L \tilde e^-_L$ &     19.4     \\
$\tilde \nu_e \tilde \nu_e$ &  25.2   \\
$\tilde e_L^\pm \tilde \nu_e$ &  80.7        \\
\hline
$\tilde \mu_R^+ \tilde \mu_R^-$ & $ 2 \cdot 10^{-5}$         \\
$\tilde \mu_L^+ \tilde \mu_L^-$ & 19.4         \\
$\tilde \nu_\mu \tilde \nu_\mu$ & 25.2         \\
$\tilde \mu^\pm \tilde \nu_\mu$ & 80.7         \\
\hline
$\tilde \chi^0_3 \tilde \chi_2^\pm$ &  $7 \cdot 10^{-3}$       \\ 
$\tilde \chi^0_4 \tilde \chi_2^\pm$ &  $7 \cdot 10^{-3}$       \\
$\tilde \chi^0_3 \tilde \chi_4^0$ &    $4 \cdot 10^{-3}$       \\
$\tilde \chi^+_2 \tilde \chi_2^-$ &    $4 \cdot 10^{-3}$       \\
\hline
\end{tabular}
\hspace{10mm}
\begin{tabular}{|rl|c|}
\hline
& mode & BR\,[\%] \\ 
\hline
$\tilde \chi_1^\pm$ & $ \to \tilde \chi_1^0 \nu_e e^\pm$ & 50 \\
                    & $ \to \tilde \chi_1^0 \nu_\mu \mu^\pm$ & 50 \\
\hline
$\tilde \chi_2^0$ & $ \to \tilde \chi_1^0 \gamma$ & 46 \\
                        & $ \to \tilde \chi_1^0 \nu_e \bar \nu_e$ & 23 \\
                    & $ \to \tilde \chi_1^0 \nu_\mu \bar \nu_\mu$ & 23 \\
                        & $ \to \tilde \chi_1^0 e^+ e^-$ & 4 \\
                        & $ \to \tilde \chi_1^0 \mu^+ \mu^-$ & 4 \\
\hline
$\tilde e_R^\pm$ & $ \to \tilde \chi_1^0 e^\pm$ & 100 \\
\hline
$\tilde e_L^\pm$ & $ \to \tilde \chi_1^\pm \nu_e$ & 58 \\
      & $ \to \tilde \chi_2^0 e^\pm$ & 29 \\
      & $ \to \tilde \chi_1^0 e^\pm$ & 12 \\
\hline
$\tilde \nu_e$ & $ \to \tilde \chi_1^\pm e^\mp$ & 55 \\
      & $ \to \tilde \chi_2^0 \nu_e$ & 28 \\
      & $ \to \tilde \chi_1^0 \nu_e$ & 17 \\
\hline
$\tilde \mu_L^\pm$ & $ \to \tilde \chi_1^\pm \nu_\mu$ & 58 \\
      & $ \to \tilde \chi_2^0 \mu^\pm$ & 29 \\
      & $ \to \tilde \chi_1^0 \mu^\pm$ & 12 \\
\hline
$\tilde \nu_\mu$ & $ \to \tilde \chi_1^\pm \mu^\mp$ & 55 \\
      & $ \to \tilde \chi_2^0 \nu_\mu$ & 28 \\
      & $ \to \tilde \chi_1^0 \nu_\mu$ & 17 \\
\hline
\end{tabular}
\caption{\small \label{tb:xs_br} \small
The NLO production cross-sections at the 13~TeV LHC and 
branching ratios of light sparticles for BP-1. 
}
\end{center}
\end{table}

There are several light sparticles which have sizeable production cross-sections
and are potentially visible.  They are the wino-like chargino and neutralino, two selectrons and sneutrinos and left-handed smuon. 
In order to avoid LHC constraints the spectrum of light sparticles must be compressed. 
The wino-like chargino and neutralino have the largest cross-section by far. 
It is worth mentioning that a pure wino LSP is disfavoured because in such a case the charged wino is mass-degenerate with the LSP and becomes long-lived leaving disappearing track signature.
The most recent disappearing track searches exclude the pure wino LSP up to $m_{\widetilde W} \sim 700$ GeV \cite{Sirunyan:2018ldc,Aaboud:2017mpt}.\footnote{This bound can be, in principle, avoided by introducing wino mixing with bino which could increase the mass splitting between wino-like chargino and neutralino which would make the disappearing track search ineffective~\cite{Badziak:2015qca}. However, in our setup this is not possible because bino-wino mixing is suppressed by both large $\mu$ and large $\tan\beta$, see e.g.~\cite{Badziak:2015qca} for approximate formulae for bino-wino mixing and the mass splitting between wino-like chargino and neutralino.}
That is why in BP-1 we choose $M_1$ and $M_2$ such that bino is the LSP, while wino is the NLSP.
\footnote{Even though $|M_1| > |M_2|$ in BP-1, in the spectrum calculated by {\tt Suspect} which includes radiative corrections~\cite{Pierce:1996zz} the lightest SUSY state is almost pure bino. }

The production process with the largest cross-section is $\tilde \chi^\pm_1 \tilde \chi^0_2$, 
which amounts to $\sigma_{\tilde \chi^\pm_1 \tilde \chi^0_2} \simeq 2.44$\,pb. 
The $\tilde \chi^0_2$ decays roughly half of the time invisibly and the other half to the LSP and a soft photon, while $\tilde \chi^\pm_1$ decays via an off-shell slepton into a charged lepton, a neutrino and the LSP. Thus in this production mode the signature is either a soft lepton + $\etmiss$ or a soft lepton + a soft photon + $\etmiss$. 
We did not find any relevant search performed by the LHC collaborations that can be sensitive to 
this signature, given the fact that 
the mass splitting between the bino and wino is only 4 GeV.

The production mode that has the next-to-the largest cross-section is 
$\tilde \chi^+_1 \tilde \chi^-_1$ with $\sigma_{\tilde \chi^+_1 \tilde \chi^-_1} \simeq 1.21$\,pb.
This cross-section is about a half of that of $\tilde \chi^\pm_1 \tilde \chi^0_2$. 
It leads to a final state with two soft leptons + $\etmiss$. 
The CMS search for two soft opposite-sign leptons based on about 35~$\rm{ fb}^{-1}$ of data have been published in ref.~\cite{Sirunyan:2018iwl}. 
In this analysis, the limits were obtained only for the chargino-LSP mass splitting above 8 GeV. 
Since the corresponding mass splitting is 4 GeV in BP-1, this limit cannot be applied. 
More recently, ATLAS presented preliminary results in similar search channel but allowing for smaller mass splittings using about 139~fb$^{-1}$ of data~\cite{ATLAS:2019lov}. For the chargino-LSP mass splitting of 4~GeV, ATLAS obtained a lower limit on the wino-like chargino mass of about 170~GeV assuming the $\tilde \chi^\pm_1 \tilde \chi^0_2 \to W^* Z^* \tilde \chi_1^0 \tilde \chi_1^0$ process. 
The BP-1 evades this constraints because $m_{\tilde \chi^\pm_1} > 170$ GeV.
Also, unlike ATLAS's assumption, the dilepton signature in BP-1 comes not from 
$\tilde \chi^\pm_1 \tilde \chi^0_2$ but from $\tilde \chi^+_1 \tilde \chi^-_1$ production,
whose cross-section is smaller than $\tilde \chi^\pm_1 \tilde \chi^0_2$. 
Moreover, it should be stressed that ATLAS has derived their constraint by performing a shape fit analysis for variables that optimises search sensitivity for a given signature. 
Since the event topologies in BP-1 are different from those in the ATLAS ones, 
one cannot directly map the ATLAS constraints to BP-1 due to different kinematics. 
For similar reasons, one cannot apply the constraint of the slepton search in the same conference note, which assumes a two-body decay of sleptons, while it is a three-body decay in BP-1.

Slepton pair production cross-section is more than one order of magnitude (even a factor of a hundred for right-handed sleptons) smaller than that of the chargino pair production. 
In BP-1, the right-handed selectron has the mass around 200~GeV and decays into a lepton and the bino-like LSP, while the left-handed sleptons decays mainly into the wino-like chargino or the neutralino accompanied by a neutrino or a lepton, respectively. 
The mass splitting between sleptons and bino-like (wino-like) states is about 30 (25)~GeV in BP-1. 
This is too small to be constrained by the standard di-slepton searches with large $\etmiss$ (see, \cite{ATLAS:2019cfv} and \cite{Sirunyan:2018nwe}). 
On the other hand, the ATLAS soft leptons search~\cite{ATLAS:2019lov} particularly targets
compressed slepton-LSP spectra. 
However, they found a limit for the mass splitting below 20~GeV, while did not set any relevant constraint for the mass splitting above 25~GeV. 
The sneutrino-slepton production cross-section is somewhat larger than the slepton pair production cross-section. However, this increase of cross-section is partially compensated by the fact that about one third of sneutrinos decay into totally invisible final states. 
Sneutrinos decay about half of the time into the chargino and a lepton. 
While there is no existing search that excludes this model point, 
it is interesting to note that a signature of the sneutrino-smuon production in this scenario is three soft leptons in the final state (or even four leptons if slepton decays to chargino rather than to neutralino).

\section{A case with the alignment from Higgs mediation}
\label{sec:Higgsmed}

While the simultaneous explanation of the $\gmte$ and $\gmtm$ anomalies presented in the previous section
is somewhat straightforward, it asks a large mass splitting between selectrons and the right-handed smuon,
$m_{\tilde E_2} \gg m_{\tilde E_1}$ (while $m_{\tilde L_2} \simeq m_{\tilde L_1}$).
In general, such a mass splitting generates large flavour-violating off-diagonal entries in the slepton mass matrices when the lepton mass matrix is diagonalised.
These off-diagonal entries are phenomenologically dangerous since it leads to 
lepton flavour violating processes such as $\mu\to e \gamma$,
which is strongly constrained experimentally~\cite{TheMEG:2016wtm}.
In order to keep this decay rate below the experimental upper limit, the slepton mass matrices must be aligned with the lepton mass matrix, so that they are simultaneously diagonalised with the common rotation matrices. 
In the scenario presented in the previous section, we must assume this alignment but did not find any good theoretical justification.

In this section, we work within a framework in which this alignment is naturally realised due to the mixed contributions 
from the Higgs mediation~\cite{Evans:2011bea,Yin:2016shg} and the usual flavour-universal contributions to the soft mass matrices.
The Higgs mediation gives the soft mass terms that are proportional to 
a product of Yukawa matrices, ${\bf Y} {\bf Y}^\dagger$ or ${\bf Y}^\dagger {\bf Y}$
depending on the handedness~\cite{Evans:2011bea,Yin:2016shg}. 
The sfermion masses in this framework can be written as
\beqn
\widetilde {\bf m}^2_L &=& m^2_L \cdot {\bf 1} + m_H^2 \cdot Y_\mu^{-2} \cdot {\bf Y} {\bf Y}^\dagger \,,
\nonumber \\
\widetilde {\bf m}^2_R &=& m^2_R \cdot {\bf 1} + m_H^2 \cdot Y_\mu^{-2} \cdot {\bf Y}^\dagger {\bf Y} \,,
\eeqn
where $m^2_L$ and $m^2_R$ are the usual flavour-universal SUSY breaking soft masses,
$Y_\mu$ is the muon Yukawa coupling 
and $m_H^2$ parametrises the size of the Higgs mediation contribution.
As can be seen, these matrices are diagonalised with the same rotation matrices as
$(\widetilde {\bf m}^2_L)^{\rm diag} = V_L^\dagger \widetilde {\bf m}^2_L V_L$,
$(\widetilde {\bf m}^2_R)^{\rm diag} = V_R^\dagger \widetilde {\bf m}^2_R V_R$
with ${\bf Y}^{\rm diag} = V_L^\dagger {\bf Y} V_R$.
Since the Higgs mediation contribution is proportional to the squared of the Yukawa coupling,
one can safely neglect this contribution to the selectron when $m_H^2 \sim m_R^2, m_L^2$.
Therefore we can parametrise our soft masses as
\begin{align}
 & m_{\tilde E_1}^2 = m_R^2 \,,   & m_{\tilde L_1}^2 = m_L^2\,,~~~~~~~~  \nonumber \\
 & m_{\tilde E_2} ^2= m_R^2 + m_H^2 \,,   & m_{\tilde L_2}^2 = m_L^2 + m_H^2 \,.
 \label{eq:Hspec}
\end{align}          
Note that this framework cannot accommodate the mass spectrum presented in the previous section 
because it is not possible to make $m_{\tilde E_2}$ heavy while 
keeping $m_{\tilde E_1}$, $m_{\tilde E_2}$, $m_{\tilde L_2}$ light.
This also implies that staus are essentially decoupled in this setup with masses of at least few~TeV.

In the Higgs mediation scenario, Eq.~\eqref{eq:Hspec}, 
the mass hierarchy is generated both in the left and right-handed sectors by $m_H^2 \gg m_L^2, m_R^2$.
This suppresses not only the bino-smuon contribution but also the chargino-sneutrino contribution to $\gmtm$. We note, however, that the bino contribution decreases with increasing smuons masses faster than the chargino contribution as long as (at least) one of the charginos is heavier than smuons. In the limit of heavy smuons (assuming bino lighter than smuons), the bino contribution scales as:
\begin{equation}
a^{\chi^0}_\mu\sim \frac{\mu M_1}{m_{\tilde{\mu}_L}^2 m_{\tilde{\mu}_R}^2} \,.
\end{equation}
Scaling of the chargino-sneutrino contribution depends on the hierarchy of $|\mu|$, $|M_2|$ and sneutrino mass. In the limit $|M_2|,|\mu|\gg m_{\tilde{\nu}_{\mu}}$, it is
\begin{equation}
a^{\chi^\pm}_\mu\sim \frac{1}{\mu M_2} \,.
\end{equation}
We see that the bino contribution scales with the common smuon mass $m_{\tilde{\mu}}$ as $1/m_{\tilde{\mu}}^4$, while the chargino contribution does not depend on $m_{\tilde{\mu}}$ up to an $\mathcal{O}(1)$ factor originating from the loop function~\eqref{loopf1}. 
However,  in the part of the parameter space where $|M_2|,|\mu|\gg m_{\tilde{\nu}_{\mu}}$, the LHC constraints on the chargino mass are very strong, especially for  wino decaying via intermediate sleptons which is excluded up to about 1.1~TeV~\cite{Sirunyan:2017lae} and SUSY contribution to muon $g-2$ cannot be large enough to explain the observed discrepancy between theory and experiment. 

If only one chargino is much heavier than sneutrino, the chargino-sneutrino contribution scales as:
\begin{equation}
a^{\chi^\pm}_\mu\sim \frac{\mu M_2}{m_{\tilde{\chi}_2^\pm}^2 m_{\tilde{\nu}_\mu}^2} \,,
\end{equation} 
where  $m_{\tilde{\chi}_2^\pm}=\max(|\mu|,|M_2|)$. 
We see that in this case the chargino-sneutrino contribution scales as $1/m^2_{\tilde{\nu_\mu}} \sim 1/m^2_{\tilde{\mu}}$, namely its decoupling for $m_{\tilde \mu} \to \infty$ is slower than
the bino contribution whose scaling is $1/m^4_{\tilde \mu}$.

\begin{figure}[t!]
\centering
\includegraphics[scale=0.67]{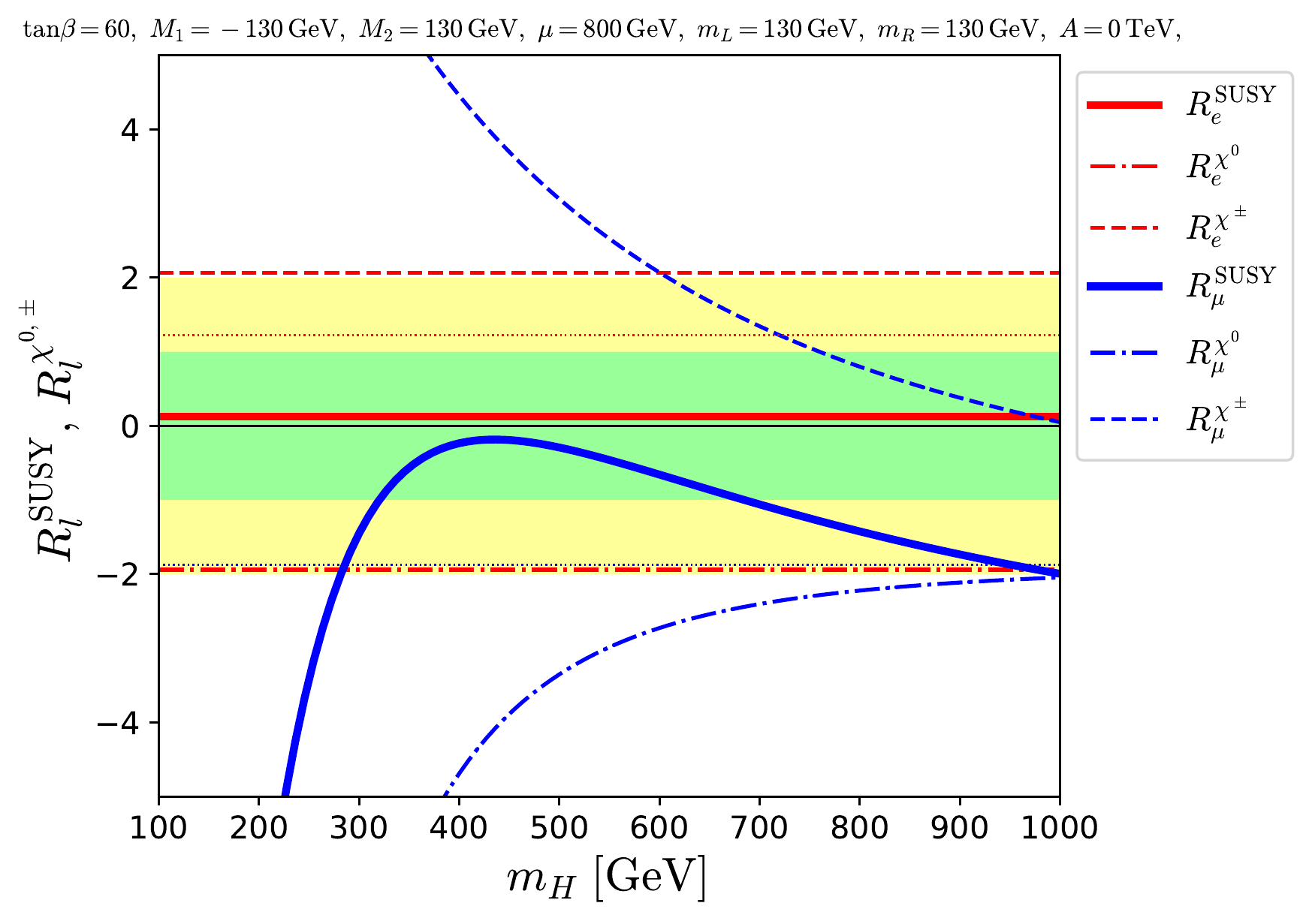} 
\caption{\small  The same as in  fig.~\ref{fig:1D_muR} but as a function of $m_H$ in the scenario motivated by the Higgs mediation of SUSY breaking. Values of the relevant SUSY parameters are given above the plot.}
\label{fig:1D_Higgsmed}
\end{figure}

In fig.~\ref{fig:1D_Higgsmed} we show $R_l$'s as a function of $m_H$. The selectron, wino and bino soft masses are fixed to 130~GeV, while the higgsino mass is set to 800~GeV, which results in the SUSY prediction of $\gmte$ in a good agreement with the experimental value.
We see that for small $m_H$
the bino contribution, which is negative, dominates $\amuSUSY$.
When $m_H$ reaches around 300 GeV, the chargino contribution starts to dominate, and for a range of $m_H$ between about 300 and 700 GeV, $\gmtm$ is in a good agreement with the experimental value 
within $1\sigma$. 
For larger $m_H$ the chargino contribution to $\amuSUSY$ is too small to fit well the experimental data.

\begin{figure}[h!]
\includegraphics[scale=0.65]{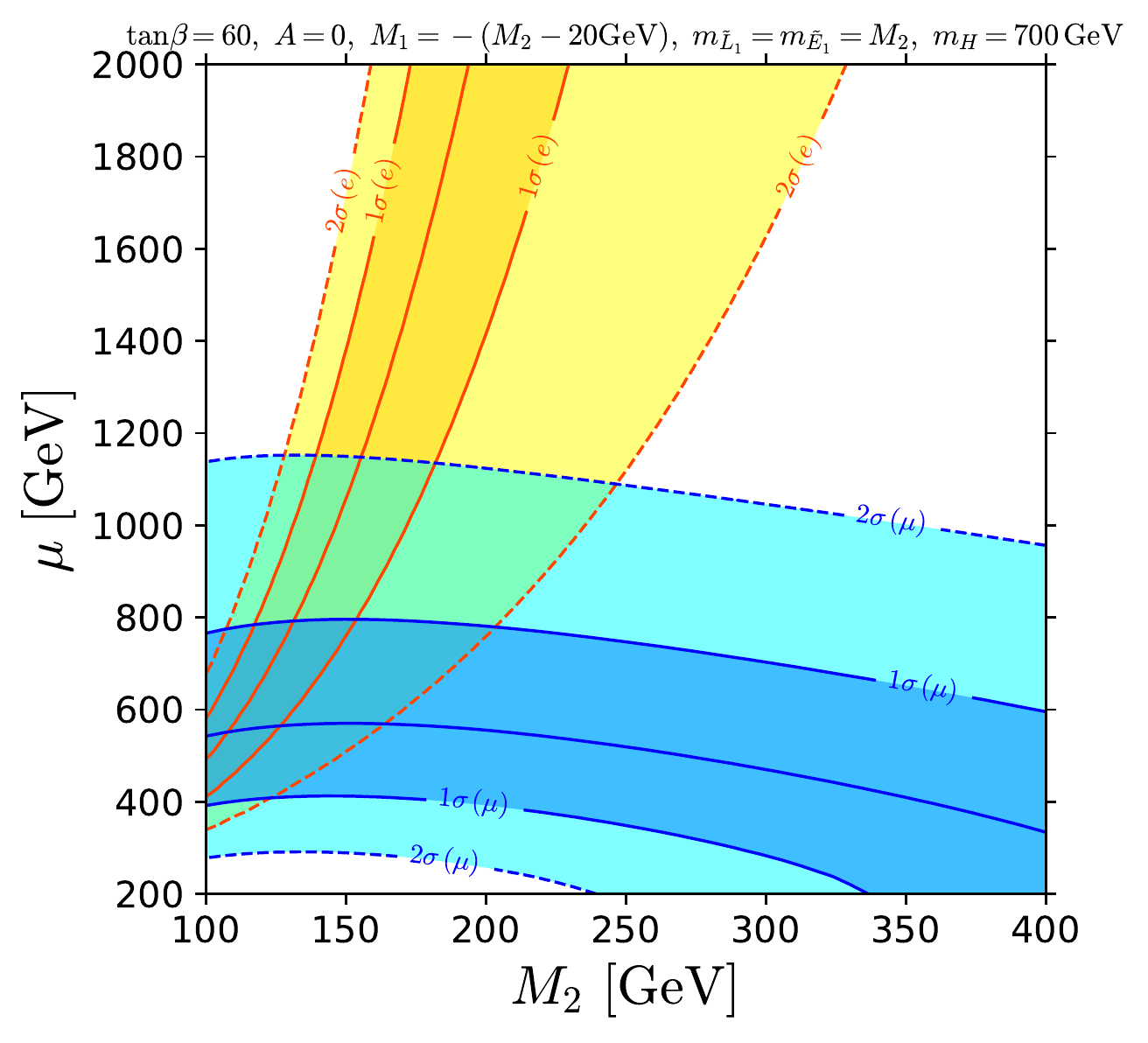}
\includegraphics[scale=0.65]{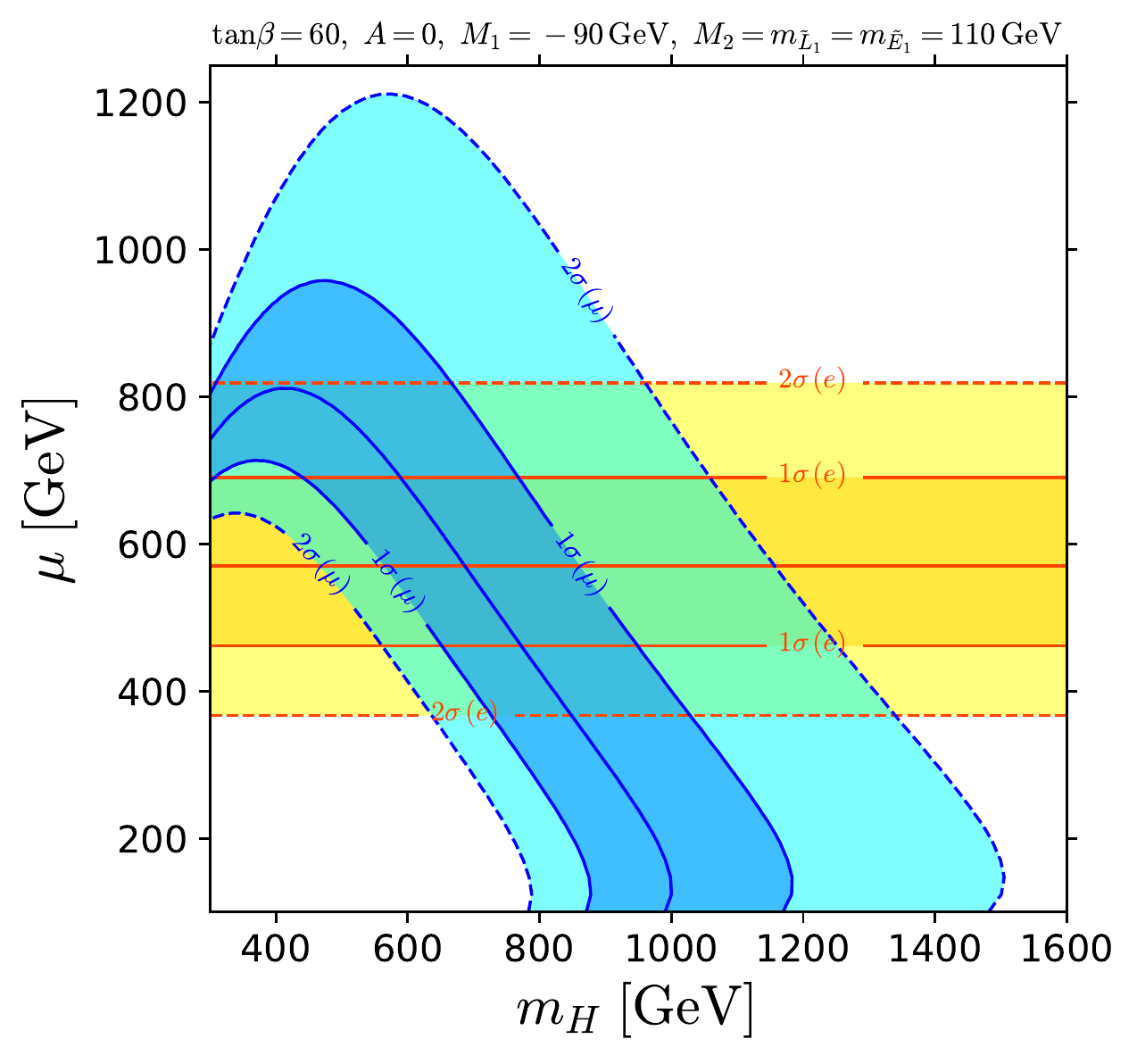}
\caption{\small
Left panel: The same as in fig.~\ref{fig:g2_mu_mL} but in the scenario motivated by the Higgs mediated SUSY breaking with $m_H=700$~GeV. Right panel: The same as in fig.~\ref{fig:g2_mu_tan} but in the $\mu$--$m_H$ plane.  }
\label{fig:2D_Higgsmed}
\end{figure}

Selectrons and wino in this scenario are even lighter than in the previous scenario with the heavy right-handed smuon, and the preferred value of $\mu$ is also smaller. 
This is because in the present case the left-handed smuon is also heavy, which suppresses the chargino contribution to $\gmtm$. 
This suppression can only be compensated by smaller $\mu$. 
On the other hand, smaller $\mu$ suppresses the bino contribution to $\gmte$, which must be compensated by lighter selectrons. 
From the left panel of fig.~\ref{fig:2D_Higgsmed} we see that for $m_H=700$~GeV, the selectron and wino masses have to be below 150 (250)~GeV to explain both $\gmtm$ and $\gmte$  at $1\sigma$ ($2\sigma$) level. 
A preferred value of $\mu$ is around $400 - 800$ GeV, which can be seen also from the right panel of fig.~\ref{fig:2D_Higgsmed}. We also see that $m_H$ has to be below 1~TeV to explain both $\gmtm$ and $\gmte$  at $1\sigma$ level.  Notice also that there is an anti-correlation between $\mu$ and $m_H$ so the upper bound on $m_H$ gets stronger for larger $\mu$. 

 Let us finally note that in this scenario values of $\tan\beta$ which are required to explain the data have to be larger than in the previous case of only the right-handed smuon being heavy. We 
 found that even for the selectron and wino masses equal to 110~GeV, i.e.~just above the LEP bound, $\tan\beta$ has to be above 40 to explain both $\gmtm$ and $\gmte$  at $1\sigma$ level.

\subsection{LHC constraints}

In order to discuss the LHC constraints let us investigate a benchmark point BP-2 defined by the following parameters: 
\beqn
M_1 = -125\,{\rm GeV},~~~M_2 = 118\,{\rm GeV},~~~
m_R = 120\,{\rm GeV},~~~
m_L = 140\,{\rm GeV},~~~ 
\nonumber \\
m_H = 700\,{\rm GeV},~~~
\mu = 700\,{\rm GeV},~~~
A = 0,~~~
\tan \beta = 60,~~~
~~~~~~~~
\eeqn
With this set of parameters, the SUSY contributions to $(g-2)_e$ and $(g-2)_\mu$ are:
\begin{align}
\aeSUSY &= -\,6.71 \cdot 10^{-13},  & R^{\rm SUSY}_e & =0.58, &   (a_e^{\tilde \chi_1^0}, a_e^{\tilde \chi_1^\pm}) &= (-10.170, ~3.459) \cdot 10^{-13},             \\
\amuSUSY &= ~~\,2.21 \cdot 10^{-9},  &  R^{\rm SUSY}_\mu & =-0.73, &
(a_\mu^{\tilde \chi_1^0}, a_\mu^{\tilde \chi_1^\pm}) &= (-0.336,~ 2.544) \cdot 10^{-9},
\end{align}
which are within the 1-$\sigma$ bands of the measurements. 
The masses of the physical states are shown in Table \ref{tb:mass2}, and the relevant branching ratios
and cross-sections are presented in Table~\ref{tb:xs_br2}.

\begin{table}[t!]
\begin{center}
\begin{tabular}{cc}
$\tilde \chi_1^0$ & 121.2 \\
$\tilde \chi_2^0$ & 123.5 \\
$\tilde \chi_1^\pm$ & 123.5 \\
\end{tabular}
~~
\begin{tabular}{cc}
$\tilde \nu_e$ & 124.7 \\
$\tilde e_L$ & 147.3 \\
$\tilde e_R$ & 127.9 \\
\end{tabular}
~~
\begin{tabular}{cc}
$\tilde \nu_\mu$ & 711 \\
$\tilde \mu_L$ & 715.3 \\
$\tilde \mu_R$ & 711.6 \\
\end{tabular}
~~
\begin{tabular}{cc}
$\tilde \chi_3^0$ & 711.9 \\
$\tilde \chi_4^0$ & 713.3 \\
$\tilde \chi_2^\pm$ & 715.7 \\
\end{tabular}
\caption{\small \label{tb:mass2} \small
Physical masses in GeV at the benchmark model point BP-2.}
\vspace{1cm}
\begin{tabular}{|r|c|}
\hline
mode & $\sigma$\,[fb] \\ 
\hline
$\tilde \chi_1^\pm \tilde \chi_2^0$ & $8.89 \cdot 10^3$ \\
$\chi_1^+ \tilde \chi_1^-$ & $ 4.48 \cdot 10^3$ \\
\hline
$\tilde e^+_R \tilde e^-_R$ &     40.9     \\
$\tilde e^+_L \tilde e^-_L$ &     66.3     \\
$\tilde \nu_e \tilde \nu_e$ &  122.6   \\
$\tilde e_L^\pm \tilde \nu_e$ &  321.2        \\
\hline
$\tilde \mu_R^+ \tilde \mu_R^-$ & 0.03         \\
$\tilde \mu_L^+ \tilde \mu_L^-$ & 0.08         \\
$\tilde \nu_\mu \tilde \nu_\mu$ & 0.07         \\
$\tilde \mu^\pm \tilde \nu_\mu$ & 0.28         \\
\hline
$\tilde \chi^0_3 \tilde \chi_2^\pm$ &  2.13       \\ 
$\tilde \chi^0_4 \tilde \chi_2^\pm$ &  2.14       \\
$\tilde \chi^0_3 \tilde \chi_4^0$ &    1.05       \\
$\tilde \chi^+_2 \tilde \chi_2^-$ &    1.15       \\
\hline
\end{tabular}
\hspace{10mm}
\begin{tabular}{|rl|c|}
\hline
& mode & BR\,[\%] \\ 
\hline
$\tilde \chi_1^\pm$ & $ \to \tilde \chi_1^0 \nu_e e^\pm$ & 100 \\
\hline
$\tilde \chi_2^0$ & $ \to \tilde \chi_1^0 \gamma$ & 12 \\
                        & $ \to \tilde \chi_1^0 \nu_e \bar \nu_e$ & 88 \\
\hline
$\tilde e_R^\pm$ & $ \to \tilde \chi_1^0 e^\pm$ & 100 \\
\hline
$\tilde e_L^\pm$ & $ \to \tilde \chi_1^\pm \nu_e$ & 59 \\
			& $ \to \tilde \chi_2^0 e^\pm$ & 30 \\
			& $ \to \tilde \chi_1^0 e^\pm$ & 11 \\
\hline
$\tilde \nu_e$ & $ \to \tilde \chi_1^\pm e^\mp$ & 36 \\
			& $ \to \tilde \chi_2^0 \nu_e$ & 18 \\
			& $ \to \tilde \chi_1^0 \nu_e$ & 47 \\
\hline
$\tilde \mu_R^\pm$ & $ \to \tilde \chi_1^0 \mu^\pm$ & 100 \\
\hline
$\tilde \mu_L^\pm$ & $ \to \tilde \chi_1^\pm \nu_\mu$ & 60 \\
			& $ \to \tilde \chi_2^0 \mu^\pm$ & 30 \\
			& $ \to \tilde \chi_1^0 \mu^\pm$ & 10 \\
\hline
$\tilde \nu_\mu$ & $ \to \tilde \chi_1^\pm \mu^\mp$ & 61 \\
			& $ \to \tilde \chi_2^0 \nu_\mu$ & 30 \\
			& $ \to \tilde \chi_1^0 \nu_\mu$ & 10 \\
\hline
\end{tabular}
\hspace{3mm}
\begin{tabular}{|rl|c|}
\hline
& mode & BR\,[\%] \\ 
\hline
$\tilde \chi_3^0$ & $ \to \tilde \chi_1^\pm W^\mp$ & 59 \\
                  & $ \to \tilde \chi_2^0 Z$ & 20 \\
                  & $ \to \tilde \chi_1^0 Z$ &  3 \\
                  & $ \to \tilde \chi_2^0 h$ &  9 \\
                  & $ \to \tilde \chi_1^0 h$ &  6 \\
\hline
$\tilde \chi_4^0$ & $ \to \tilde \chi_1^\pm W^\mp$ & 59 \\
                  & $ \to \tilde \chi_2^0 h$ & 20 \\
                  & $ \to \tilde \chi_1^0 h$ &  3 \\
                  & $ \to \tilde \chi_2^0 Z$ & 10 \\
                  & $ \to \tilde \chi_1^0 Z$ &  6 \\
\hline
$\tilde \chi_2^\pm$ & $ \to \tilde \chi_1^\pm Z$ & 30 \\
                    & $ \to \tilde \chi_1^\pm h$ & 29 \\  
                    & $ \to \tilde \chi_2^0 W^\pm$ & 29 \\
                    & $ \to \tilde \chi_1^0 W^\pm$ & 9 \\
                    & $ \to \tilde e^\pm \nu_e$ & 1.5 \\
\hline
\end{tabular}
\caption{\small \label{tb:xs_br2} \small
The NLO production cross-sections at the 13~TeV LHC and 
branching ratios of light sparticles for BP-2.}
\end{center}
\end{table}

We see that in order to explain both $\gmte$ and $\gmtm$ the masses of selectrons and wino-like states are smaller than in BP-1. 
The corresponding production cross-sections are larger by a factor of five. 
Therefore, in order to avoid LHC constraints we expect to need even smaller mass splittings. 
We see from Table~\ref{tb:xs_br2} that the splitting between the wino-like chargino and the neutralino is about 2 GeV. For such splitting the ATLAS soft dilepton search~\cite{ATLAS:2019lov} sets a lower bound on chargino mass of about 100~GeV i.e.~comparable to the LEP bound. 
Moreover, similarly to BP-1 the ATLAS search cannot be applied directly due to different kinematics. 
It is noteworthy that in the present case decays of the wino-like chargino result only in electrons in the final state since decays to muons are suppressed by a large smuon mass. 
Another difference from BP-1 is that decays of the wino-like neutralino are mostly invisible, which is due to a smaller mass splitting between the electron-type sneutrino and the LSP.  

In contrast to BP-1, smuons in the present case are neither compressed nor decoupled, so the LHC constraints are in principle relevant. 
For 100~GeV LSP, CMS set a lower bound on the smuon mass, which is about 300~GeV by a search for the direct pair production of degenerate left- and right-handed smuons using 35.9 fb$^{-1}$ of data~\cite{Sirunyan:2018nwe}. 
More recent ATLAS preliminary results based on 139 fb$^{-1}$ of data provides a lower bound of about 700~GeV, but assuming full mass degeneracy between the first and the second generation of sleptons~\cite{ATLAS:2019cfv}. While the ATLAS search cannot be used directly to set the limit on smuons one can approximately use it in our scenario noting that the pair production cross-section for the first two generations of 700~GeV sleptons is about 0.24~fb, which can be used as the cross-section upper limit.  
In BP-2, the cross-section for pair production of right-handed and left-handed smuons is 0.11~fb, which is smaller than this limit. 
One should also take into account that left-handed smuons decay into muons and $\tilde \chi^0_1$ or $\tilde \chi^0_2$ with branching ratios of about 40~\% and that dimuons in final state result also from the pair production of muon-type sneutrinos and sneutrino-smuon associated production. 
The total dimuon production cross-section from all of those modes is about 0.14~fb, which is still smaller than the corresponding ATLAS limit 0.24~fb.

Another difference between BP-1 and BP-2 is the higgsino mass. 
While $\mu = 1700$ GeV in BP-1, it is 700 GeV in BP-2.
Despite the lighter higgsinos, the BP-2 still satisfies the LHC constraints.
Generally, the limit on the chargino mass is the strongest if it decays into the LSP via on-shell sleptons.
The limit in this case is about 1 TeV.
On the other hand, this decay mode is strongly suppressed in BP-2 with the branching ratio of only $\sim 1$~\%. 
The higgsinos in BP-2
decay mainly to the LSP via on-shell bosons, $W^\pm$, $Z^0$ and $h$, instead. 
The strongest constraint in this type of topologies is given by the recently reported ATLAS search targeting the $pp \to \tilde \chi_1^\pm \tilde \chi_2^0 \to W^\pm h \tilde \chi_1^0 \tilde \chi_1^0$ topology~\cite{ATLAS:2019efx}. In that analysis a lower bound on the chargino mass was set to $\sim 700$~GeV for the LSP mass of 100~GeV with the assumption that $\tilde \chi_1^\pm$ and $\tilde \chi_2^0$
are wino-like. 
We estimate that this corresponds to a lower bound on higgsino-like chargino mass of about 600~GeV since the higgsino production cross-section is smaller by a factor of two compared to the wino case with the same mass. 
There is also a relevant CMS analysis targeting the $pp \to \tilde \chi_1^\pm \tilde \chi_2^0 \to W^\pm Z \tilde \chi_1^0 \tilde \chi_1^0$ topology \cite{Sirunyan:2018ubx}.
This search places an upper bound on the chargino-neutralino production of $\sim 10$~fb, which is weaker than the total higgsino production cross-section including all production modes, which amounts to $\sim 6$~fb, as seen from Table~\ref{tb:xs_br2}. 
Moreover, all those searches assume that the chargino and neutralino have only one decay channel, while in BP-2 several decay channels are equally important so the lower limit on chargino mass is expected to be much weaker than those provided by the experimental analyses in simplified models.

\section{Conclusions}
\label{sec:concl}

We have investigated the scenarios in the MSSM that explains both electron and muon $g-2$ anomalies.
It has been demonstrated that this is possible without introducing explicit flavour-mixing
by arranging the bino-slepton and chargino-sneutrino contributions differently between
the electron and muon sectors. 
We identified features of the sparticle spectrum where smuons are not degenerate with selectrons and the sign of $M_1M_2$ is negative. Moreover, selectrons and wino must be very light, while higgsino tends to be heavier with mass $\mathcal{O}(1)$~TeV and large $\tan\beta$ is preferred.

We analysed in detail two scenarios with a different pattern of smuon masses. 
In the scenario with the right-handed smuon much heavier than the other sleptons,  selectrons, the left-handed smuon and wino must be lighter than $\sim 200$~GeV for $\tan\beta=60$ to simultaneously fit $\gmte$ and $\gmtm$ at 1$\sigma$ level. 
In the second scenario, motivated by the Higgs mediation where the $\mu \to e \gamma$ is naturally suppressed, the left- and right-handed smuons are almost degenerate. 
Fitting $\gmte$ and $\gmtm$ at 1$\sigma$ level requires even lighter selectrons and wino with masses below $\sim 150$~GeV for $\tan\beta=60$. 
While selectrons and wino are very light, all LHC constraints can be satisfied due to the small mass splitting between wino-like chargino and neutralino, selectrons and the bino-like LSP.

\section*{Note Added}
During completion of this work Ref.~\cite{Endo:2019bcj} was submitted to arXiv where simultaneous explanation of the $\gmte$ and $\gmtm$ anomalies was also proposed in the framework of the MSSM without introducing lepton flavour mixing. However, the parameter space considered in Ref.~\cite{Endo:2019bcj} is very different than ours since they consider $\mu$ in a range of hundreds of TeV which results in very large threshold corrections to electron and muon Yukawa couplings. Predictions for the sparticle spectrum are also very different than in our setup since masses of selectrons are in the multi-TeV range.

\section*{Acknowledgments}
MB would like to thank Robert Ziegler for useful discussions. This work has been partially supported by National Science Centre, Poland, under research grant no. 2017/26/D/ST2/00225.
The work of KS is partially supported by 
the National Science Centre, Poland, under research grants 2017/26/E/ST2/00135 and
the Beethoven grants DEC-2016/23/G/ST2/04301.

\end{document}